# Frequency offset tolerant synchronization signal design in NB-IoT

Jun Zou

**Abstract – Timing detection is the first step and very important in wireless communication systems. Timing detection performance is usually affected by the frequency offset. Therefore, it is a challenge to design the synchronization signal in massive narrowband Internet of Things (NB-IoT) scenarios where the frequency offset is usually large due to the low cost requirement. In this paper, we firstly proposed a new general synchronization signal structure with a couple of sequences which are conjugated to remove the potential timing error arose from large frequency offset. Then, we analyze the suitable sequence for our proposed synchronization signal structure and discuss a special ZC sequence as an example. Finally, the simulation results demonstrate our proposed synchronization signal can work well when the frequency offset is large. It means that our proposed synchronization signal design is very suitable for the massive NB-IoT.**

## I. Introduction

INTERNET of Things (IoT) is a network that aims to connect all the devices in the world and makes the information exchange between them easily. With the rapid development of wireless communication technologies, the IoT is realized step by step in recent years. Over 60% of the IoT application scenarios are massive narrowband IoT (NB-IoT), whose characteristics include massive access devices, transmission delay tolerance and low cost, such as smart meters, eHealth and so on [1]-[3]. The low cost requirement means that cheap/inaccurate crystal oscillators are used which leads to a large frequency offset between the transmitter and the receiver. It poses challenges to the design of the downlink synchronization signal which is used to obtain the downlink timing. In tradition LTE system, the primary synchronization signal (PSS) and secondary synchronization signal (SSS) are used to achieve the downlink timing [4]. PSS detection, as the very first process, usually suffers from the large frequency offset. The frequency offset will destroy the 'perfect' correlation of the ZC sequence used by the PSS which could lead to timing error even there is no noise [4]. Then differential correlator or partial correlator is usually used to remove or diminish the effects of frequency offset on timing detection [5].

In this letter, we firstly proposed a new general synchronization signal structure with a couple of sequences which are conjugated to remove the potential timing error caused by the frequency offset. Then we analyze the suitable sequence for our proposed structure and analyze a special ZC sequence as an example. Finally, simulation results are given to verify our proposed design.

## II. SYSTEM MODEL

Assuming the synchronization signal in time domain is $x[n]$, $n = 0,1,\cdots,N-1$ at the transmitter, where $N$ is the length of the synchronization sequence. Assuming there is one receiving antenna, the received signal at the device can be written as

$$y[n] = hx[n-d]e^{j2\pi n\Delta f/f_s} + w[n],\qquad(1)$$

with the Nyquist sampling rate $f_s$ without the considering of the sampling offset, $\Delta f$ is the frequency offset between the transmitter (base station) and the receiver (device), where $h \sim CN(0,\sigma_h^2)$ is the channel gain, unknown but assumed to be constant over the PSS transmission duration, $d$ is the number of samples corresponding to the signal propagation time between the transmitter and the receiver, $w[n] \sim CN(0,\sigma_w^2)$ is the white Gaussian noise.

At receiver, a local sequence which is the copy of the transmit synchronization sequence is used to detect the right receiving start time by calculating the correlation between the received signal and the local sequence. The output of direct correlator can be written as

$$z(k) = \frac{1}{N}\sum_{n=0}^{N-1} y[n+k]x^*[n].\qquad(2)$$

When the phase change caused by the frequency offset is large in the signal duration, e.g., $\frac{2\pi\Delta fN}{f_s} > \pi$, M-part correlator or differential correlator will be used to replace the direct

correlator [5]. The output of the M-part correlator can be written as

$$z(k) = \frac{1}{N} \sqrt{\sum_{m=0}^{M-1} \left| \sum_{n=\frac{N}{M}m}^{\frac{N}{M}(m+1)-1} y[n+k]x^*[n] \right|^2} \ .$$

(3)

Obviously, the processing gain will decrease about $10\log_{10}M$ dB for the M-part correlator due to the non-coherent combination among the different parts comparing to the direct correlator.

The output of the differential correlator can be expressed as

$$z(k) = \frac{1}{N-k_{\mathrm{d}}} \sum_{n=0}^{N-1-k_{\mathrm{d}}} y[n+k_{\mathrm{d}}+k]y^*[n+k]x^*[n+k_{\mathrm{d}}]x[n],$$

(4)

where $k_{\mathrm{d}}$ is the distance between the two samples doing the differentiation. Although the differentiation operation can remove the frequency offset perfectly, there will be at least 3 dB loss on signal to noise ratio (SNR) due to the fact that the noise is amplified by the multiplication between two received samples [3]. Therefore, these two methods are usually used when the frequency offset is large.

Then a maximum likelihood estimate (MLE) is used to estimate the right timing and it can be expressed as

$$\hat{k} = \arg \max_k \left| z(k) \right|.$$

(5)

### III. CONJUGATED-SEQUENCES-BASED TIMING STRUCTURE

Our proposed synchronization signal consists of a couple of sequences which are conjugated. When we and the channel gain (constant in one detection process), the output of the direct correlator in (2) can be rewritten as

$$\left| z_1(\Delta k_1, \Delta f) \right| = \left| \frac{1}{N} \sum_{n=0}^{N-1} x[n+k_1-d]e^{j2\pi(n+k)\Delta f/f_s} x^*[n] \right| = \frac{1}{N} \left| \sum_{n=0}^{N-1} x[n+\Delta k_1]x^*[n]e^{j2\pi n\Delta f/f_s} \right|,$$

(6)

where $\Delta k_1 \triangleq k_1 - d$, $k_1$ is the index of the direct correlator output with local sequence $x[n]$.

The other synchronization sequence in our design is $x^*[n]$, $n = 0,1,\cdots,N-1$, the correlator output can be expressed as

$$\left| z_2(\Delta k_2, \Delta f) \right| = \left| \frac{1}{N} \sum_{n=0}^{N-1} x^*[n+k_2-d]x[n]e^{j2\pi n\Delta f/f_s} \right| = \left| \frac{1}{N} \sum_{n=0}^{N-1} x[n]x^*[n-\Delta k_2]e^{j2\pi n\Delta f/f_s} \right|, \qquad (7)$$

where $\Delta k_2 \triangleq d - k_2$, $k_2$ is the index of the direct correlator output with local sequence $x^*[n]$.

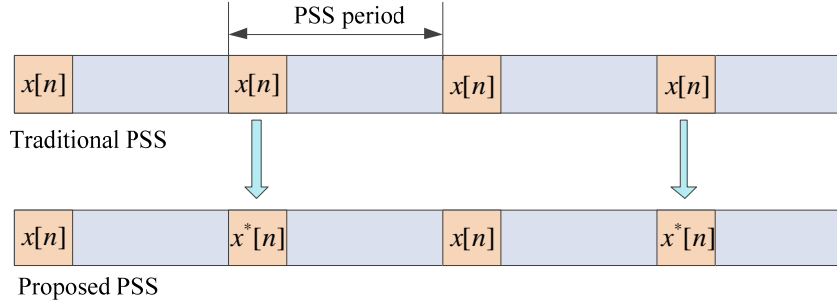

Fig. 1 Illustration of our proposed conjugated-sequences-based PSS structure.

Obviously, when $\Delta k_1 = \Delta k_2$

$$\left| z_1(\Delta k_1, \Delta f) \right| = \left| z_2(\Delta k_2, \Delta f) \right|, \qquad (8)$$

without considering the data adjacent to the synchronization sequence, that is

$$x[n] = 0, n \geq N \text{ or } n < 0 \quad . \qquad (9)$$

It means if the timing estimated from the correlator with local sequence $x[n]$ is

$$\hat{k}_1 = \Delta \hat{k}_1 + d, \qquad (10)$$

the timing estimated from the correlator with local sequence $x^*[n]$ must happen at

$$\hat{k}_2 = d - \Delta \hat{k}_2 = d - \Delta \hat{k}_1, \qquad (11)$$

where $\Delta \hat{k}_1$ and $\Delta \hat{k}_2$ are the timing estimation errors which are the same according (8).

Therefore, the right timing can be estimated according to

$$\hat{k} = \frac{\hat{k}_1 + \hat{k}_2}{2}, \qquad (12)$$

which can remove the potential reducible timing error caused by the large frequency offset which is hard to avoid [6].

In current LTE system, the synchronization signal is periodic, so we let the two PSS sequences in adjacent period utilize the $x[n]$ and $x^*[n]$ respectively as shown in Fig. 1. When the timing error caused by the frequency offset is small (for the device with expensive crystal oscillator), it just needs search one of the PSS signal rather than two for the purpose of quick search, low complexity and compatibility. Besides, it can maintain the PAPR property of the original sequence $x[n]$ comparing to add them together at one period.

Obviously, the key of our proposed design is to find a sequence whose maximum correlator output is insensitive to the frequency offset.

## IV. FREQUENCY OFFSET TOLERANT SIGNAL SELECTION

As we known, chirp signal is widely used in radar system due to the fact that its correlator output is insensitive to the frequency offset[7]. The general expression of chirp signal after sampling can be written as

$$x[n] = e^{j\pi(a_2 n^2 + a_1 n + a_0)}, \tag{13}$$

where $a_0, a_1, a_2 \in R$ are the coefficients. Then substituting (13) into (6) yields

$$\left|z_1(\Delta k_1, \Delta\lambda, a_2)\right| = \frac{1}{N}\left|\frac{\sin\left(\dfrac{\pi\Delta\lambda(N - \left|\Delta k_1\right|)}{N} + \pi a_2 \Delta k_1 (N - \left|\Delta k_1\right|)\right)}{\sin\left(\dfrac{\pi\Delta\lambda}{N} + \pi a_2 \Delta k_1\right)}\right|, \tag{14}$$

where $\Delta\lambda = \Delta f / \Delta f_s$ is the frequency offset normalized to the subcarrier spacing $\Delta f_s$. According to the property of sine function, (14) can be rewritten as

$$\left|z_1(\Delta k_1, \Delta\lambda, a_2)\right| = \frac{1}{N}\left|\frac{\sin\left(\dfrac{\pi\Delta\lambda(N - \left|\Delta k_1\right|)}{N} + \pi\xi\Delta k_1 (N - \left|\Delta k_1\right|)\right)}{\sin\left(\dfrac{\pi\Delta\lambda}{N} + \pi\xi\Delta k_1\right)}\right|, \tag{15}$$

where $\xi = a_2 - \lfloor a_2 \rfloor$, $\lfloor \cdot \rfloor$ is the integer-valued function. It means the correlator output only depends on the fractional part of coefficient $a_2$ corresponding to the sequence. For ease of discussion, we can only consider the value of $a_2 \in [0,1)$.

According to the above analysis, ZC sequence is one of the sequences satisfy the constraints and suitable for our proposed PSS design, which can be expressed as [8]

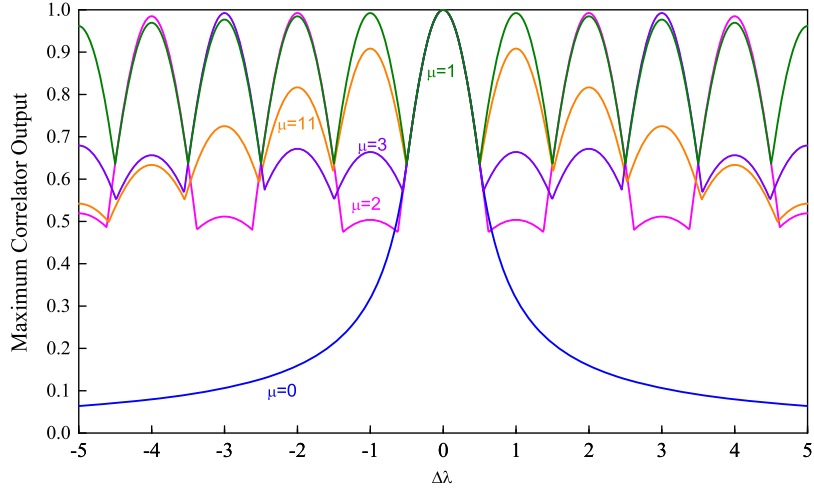

Fig. 2 Illustration of the maximum correlator output of ZC sequences with different roots ($N$=131).

$$x[n] = e^{-j\frac{\pi\mu n(n+1)}{N}}, \quad n = 0,1,2,\cdots N-1,$$ (16)

where $N$ is the length of the ZC sequence, $\mu = 0,1,2,\cdots,N-1$ is the root of the ZC sequence and $a_2 = \frac{\mu}{N}$.

Apparently, the correlator outputs of the ZC sequences with different roots are different. And the low reduction of maximum output of the correlator under large frequency offsets is benefit to our proposed method. Fig. 2 shows an example of the maximum correlator output of different ZC sequences (e.g., $N = 131$) under different frequency offsets. The maximum output of the correlator is defined as

$$z_{\max}(\Delta\lambda, a_2) = \max_{\Delta_{k_1}} \left| z_1(\Delta k_1, \Delta\lambda, a_2) \right|.$$ (17)

From Fig. 2, we can see that the ZC sequence with root $\mu = 1$ is most insensitive to the frequency offset. When $\mu = 1$, (15) can be rewritten as

$$\left| z_1(\Delta k_1, \Delta\lambda, \frac{1}{N}) \right| = \begin{cases} \dfrac{1}{N}\left(N - \left|\Delta\lambda\right|\right) & \Delta k_1 = -\Delta\lambda \\[4mm] \dfrac{1}{N}\left| \dfrac{\sin\left(\dfrac{\pi(\Delta\lambda + \Delta k_1)(N - \left|\Delta k_1\right|)}{N}\right)}{\sin\left(\dfrac{\pi(\Delta\lambda + \Delta k_1)}{N}\right)} \right| & others \end{cases}.$$ (18)

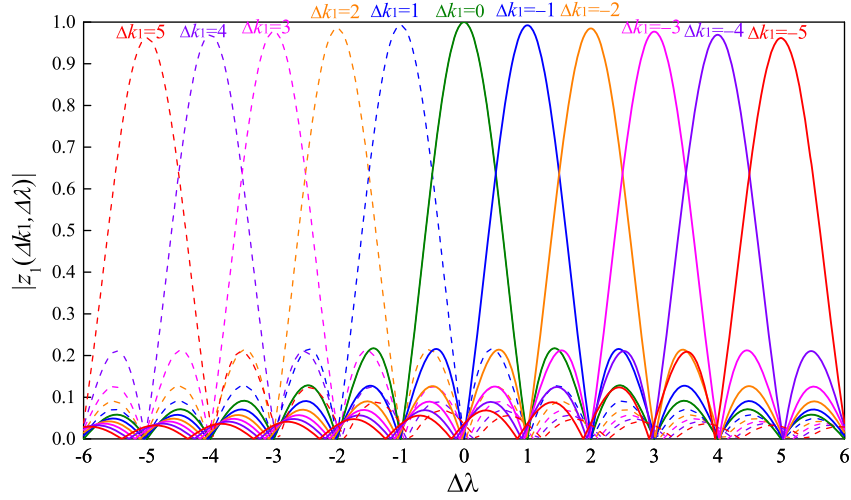

Fig. 3 Illustration of the correlator output of ZC sequence with root $\mu = 1$ under different timing offsets.

Fig. 3 shows the correlator output in (18) under different frequency offsets and different timing offsets with $\mu = 1$. We can see that there is no timing error when the frequency offset is small (e.g., $\Delta\lambda < 0.5$). However, when the frequency offset increases, the timing error is presence according to the criterion defined in (5). But the timing error can be removed easily by our proposed design. Besides, by comparing Fig. 2 and Fig. 3, we can see that the minimum of the maximum correlator output decreases slowly when the frequency offset $\Delta\lambda$ increases step by 1. Because the gap of the timing offset $\Delta k_1$ corresponding to the adjacent peaks is one which is always true with $\mu = 1$ for arbitrary $N$. Therefore, ZC sequence with $\mu = 1$ is suit for our proposed conjugated-sequences-based timing structure.

According (18) and Fig. 3, we can know that the minimum of the maximum correlator output locates at the cross point corresponding to the adjacent timing offset. When $\Delta k_1 \geq 0$, $\Delta\lambda \in [-\Delta k_1 - \frac{1}{2}, -\Delta k_1 + \frac{1}{2}]$, it is easy to show that

$$\left| z_1(\Delta k_1, \Delta\lambda, \frac{1}{N}) \right| > \left| z_1(\Delta k_1 + 1, \Delta\lambda - 1, \frac{1}{N}) \right|, \tag{19}$$

according to (18). And when $\Delta k_1 \leq 0$, $\Delta\lambda \in [\Delta k_1 - \frac{1}{2}, \Delta k_1 + \frac{1}{2}]$, it is also easy to show that

$$\left| z_1(\Delta k_1, \Delta\lambda, \frac{1}{N}) \right| > \left| z_1(\Delta k_1 - 1, \Delta\lambda + 1, \frac{1}{N}) \right|. \tag{20}$$

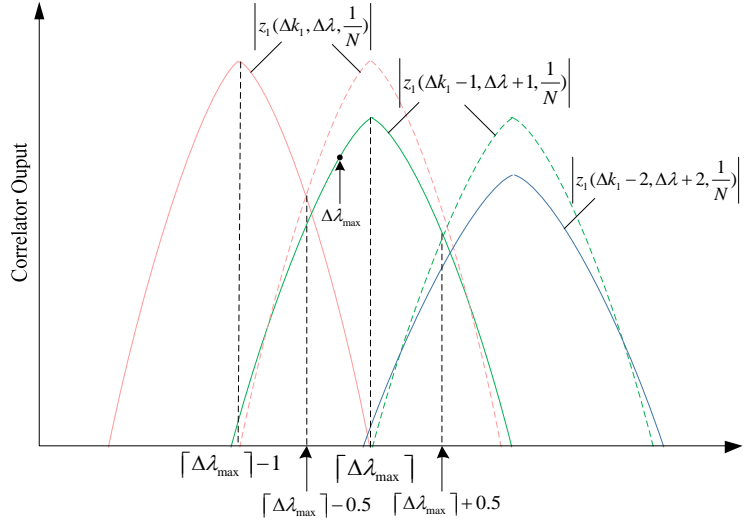

Fig. 4 Illustration of the main lobes of the correlator output around the maximum frequency offset $\Delta\lambda_{max}$ ( $\Delta\lambda_{max} > 0$ in this example). And the dash line is the copy of the solid one with the same color.

According to (19), (20) and Fig. 4, we can get that at the given maximum frequency offset $\Delta\lambda_{max}$, the minimum of the maximum correlator output satisfies

$$z_{min} > \left| z_1\left(-\lceil|\Delta\lambda_{max}|\rceil, \lceil|\Delta\lambda_{max}|\rceil + 0.5, \frac{1}{N}\right) \right| = \frac{1}{N}\left| \frac{\sin\left(\frac{\pi(N-\lceil|\Delta\lambda_{max}|\rceil)}{2N}\right)}{\sin\left(\frac{\pi}{2N}\right)} \right|. \tag{21}$$

The current narrowband PSS (NPSS) in NB-IoT is about 0.8 ms in time domain [9], so the subcarrier spacing in our proposed PSS is 1.25 kHz, and $N = 131$ for the justice comparison. When the maximum frequency offset is 40 kHz which is equal to 20 ppm at the carrier frequency 2 GHz, the maximum detection energy loss is about

$$10\log_{10}\frac{\sin\left(\frac{\pi(N-\lceil|\Delta\lambda_{max}|\rceil)}{2N}\right)}{N\sin\left(\frac{\pi}{2N}\right)} \approx -2.3 \ dB, \tag{22}$$

which is smaller than the energy loss in the differential correlator and M-part correlator.

Therefore, the ZC sequence with root $\mu = 1$ is very suitable for our proposed conjugated-sequences-based PSS structure whose correlator output is insensitive to frequency offset.

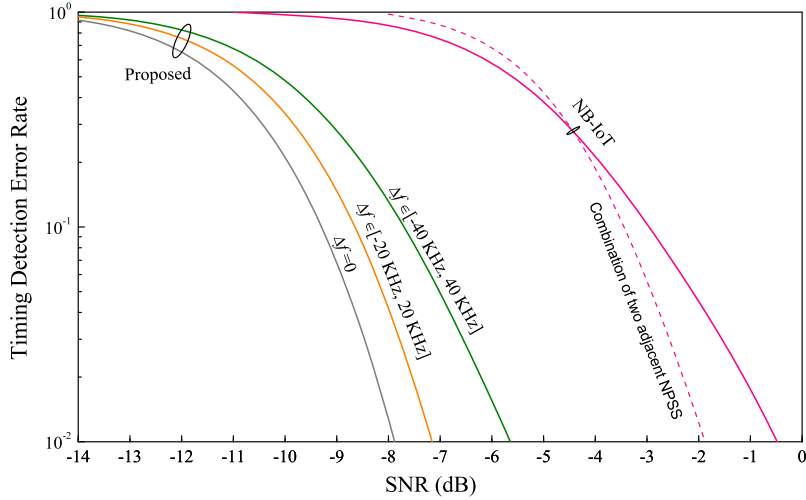

Fig. 5 Timing detection error rate of our proposed synchronization signal with different frequency offsets

## V. SIMULATION RESULTS

In this section, simulation results are given to verify the performance of our proposed PSS signal with a couple of ZC sequences which are conjugated. The system parameters are set as follows: the proposed PSS is generated in time domain in OFDM system, the subcarrier spacing is $\Delta f_s = 1.25$ kHz, the length of ZC sequence is $N = 131$, the root of ZC sequence is $\mu = 1$, the synchronization signal period is 10 ms (the same as the NPSS in NB-IoT [9]). In the simulation, the frequency offset is randomly and uniformly selected and added. AWGN channel is used in the simulations. When the timing error is larger than 1 us, we regard it as detection error. Fig. 5 shows the timing detection error rate at different SNRs. We also add the NB-IoT NPSS detection performance as a reference where the differential correlator is used. For fair comparison, we also simulated the timing performance of the combination of two adjacent NPSS that the estimated timing is the average of the timing estimated from the two adjacent NPSS. We can see that the deterioration of timing performance caused by the frequency offset is small thanks for that the correlator output of ZC sequence with root 1 is insensitive to the frequency offset. We can see that our proposed synchronization signal can work better than the NPSS with differential correlator even under the maximum frequency offset 40 kHz. It is true that the detection performance of our proposed synchronization signal will be worse than the NB-IoT PSS with the increment of frequency offset due to the fact that differentiation operation can remove the frequency offset effect on detection. But our proposed PSS design is good enough to cover the potential frequency offset range in NB-IoT scenarios (less than 40 kHz [10]).

## VI. Conclusion

This letter investigates the downlink synchronization signal design in large frequency offset scenario. We propose a general synchronization signal structure design with a couple of sequences which are conjugated to deal with the timing issue instead of the differentiation operation or partial correlator. Moreover, we discuss the general formulae of the suitable sequence for our proposed structure and choose the ZC sequence with root 1 as an example. The loss of the detection energy is small thanks for the special ZC sequence's insensitivity to frequency offset. The simulation results demonstrate our proposed synchronization signal can work better than the NPSS under large frequency offset. Therefore, our proposed synchronization signal is very suit for the low cost NB-IoT scenarios.